\documentclass[a4paper]{article}%

\usepackage{amsmath}%
\usepackage{amsfonts}%
\usepackage{amssymb}%
\usepackage{graphicx}

\usepackage{mathrsfs}
\usepackage{xspace}
\usepackage{mathtools, amssymb}
\let\curv\mathscr
\usepackage[affil-it]{authblk}
\usepackage{geometry}
\begin{document}

\title{Modified Gravity Models
\\  Admitting Second Order Equations of Motion}

\author{Aimeric Coll\'{e}aux $^{1,}$   \thanks{Electronic address: \texttt{aimericcx@yahoo.fr}} , Sergio Zerbini $^{2,}$}

\affil{
$^{1}$ Department of Physics, Trento University, Via Sommarive 14-38123 Trento, Italy \\
$^{2}$ Department of Physics, Trento University and
TIPFA-INFN, Via Sommarive 14-38123 Trento, Italy\\}

\maketitle

\textbf{Abstract} : The aim of this paper is to find higher order geometrical corrections to the Einstein-Hilbert action that can lead to only second order equations of motion.  The metric formalism is used, and static spherically symmetric and Friedmann-Lema\^{i}tre space-times are considered, in four dimensions. The FKWC-basis are introduced in order to consider all the possible invariant scalars, and  both polynomial and non-polynomial gravities are investigated.
\\


\section{Introduction}

Most of the equations of motion describing  physical effects are second order, that is, we need to specify either an initial and a final position in space-time to describe the dynamics between them, or we need an initial position and velocity to describe how the system will evolve. Concerning General Relativity (GR), for which the gravitational field $g_{\mu\nu}(x)$ is encoded into the geometry of space-time :
\begin{eqnarray}
ds^2 = g_{\mu\nu}(x) dx^{\mu}dx^{\nu} \,,
\end{eqnarray} 
The Einstein field equations, describing the dynamics of the geometry, are also second order ones. 

However, it is well known that two of the simpliest solutions of GR, the Schwarzschild metric and the Friedmann-Lema\^{i}tre one, suffer from the existence of singularities. When one is dealing with ordinary matter, this is a general fact. Furthermore this theory alone is not able to describe dark energy, even though the inclusion of a suitable cosmological constant is sufficient. But then, other problems arise, like the cosmological constant one \cite{1} and the coincidence problem.   Therefore, one can think about modifying the Einstein equations, in the hope to describe dark energy and to cure singularities. In order to do so,  one can add higher order invariant scalars, like $R_{\mu\nu}R^{\mu\nu}$, to the Einstein-Hilbert action to have high energy corrections that could describe what really happen around the singularities \cite{2,3}. With regard to dark energy issue, see
\cite{4,5,6,7,8}.

Within an higher order modified gravity model, the equations of motion will no longer be second order ones : there will be more than two initial conditions to specify in order to find the dynamics, so to keep the physical sense of what is an equation of motion, it is needed to introduce new fields for which these additional initial conditions would apply, such that at the end, the theory would involve two dynamical fields, with second order equations of motion for both of them. By doing so, we face an important problem, that is the presence of Ostrogradsky instabilities (see, for example \cite{4}) : the new field defined in this way can carry negative kinetic energy such that the Hamiltonian of the theory is not bounded from below and can reach arbitrarily negative energies, what would make this theory impossible to quantize in a satisfying way \cite{4}. And there are no general rules to avoid this problem, although a well known class of modified gravity, equivalent to GR plus a scalar field, the $f \big( R \big)$ one, might not suffer from this problem \cite{9}.

Moreover, with a new field involved in the dynamics of gravity, this last would not be a fully geometrical theory anymore, which is yet one of the most important implications of General Relativity. Nevertheless, it is possible to find second order equations of motion from the addition into the Einstein-Hilbert action of higher order scalars \cite{10}. In this way, the Ostrogradsky instability may be avoided and there are no additional field involved in the dynamics, so these corrections can be said to be "geometrical" ones. This kind of modifications are the Lovelock scalars, but it turns out that in four dimensions, the only higher order scalar made of contractions of curvature tensors (only) that leads to second order equations \cite{10} is the so-called Gauss-Bonnet invariant :
\begin{eqnarray}
\mathcal{E}_4 = R^2 - 4 R_{\alpha \beta} R^{\alpha \beta}  + R_{\alpha\beta\gamma}^{\ \ \ \ \delta}R^{\alpha\beta\gamma}_{\ \ \ \ \delta}	,
\end{eqnarray}
 which is however a total derivative in four dimensions \cite{11}, and then it does not contribute to the equation of motion:
 \begin{eqnarray}
\sqrt{-g}\mathcal{E}_4 = \partial_{\alpha} \Bigg( -\sqrt{-g} \, \epsilon^{\alpha \beta \gamma \delta} \; \epsilon_{\rho \sigma}^{\ \ \mu \nu} \Gamma_{\mu \beta}^{\ \ \ \rho} \Big( \frac{1}{2} R_{\delta\gamma\nu}^{\ \ \ \ \sigma} - \frac{1}{3} \Gamma_{\lambda \gamma}^{\ \ \ \sigma}\Gamma_{\nu \delta}^{\ \ \ \lambda} \Big) \Bigg).
\end{eqnarray}

This result is background independent, which means that if we want to find a second order correction for all possible metrics in four dimensions, then this unique term does not contribute to the dynamics.   
That is why, in order to find anyway significant corrections to General Relativity that could cure some of its problems, we will search for additional terms that will give second order equations for only some specific metrics : the most studied ones, that suffer from singularities, the FLRW space-time describing the large scale dynamics of the universe, and the static spherically symmetric space-time describing neutral non-rotating stars and black holes. 

We note however that our way to find second order corrections is not at all the only possible one. There are other formulations of GR  than the metric one, where the equations of motion are found by varying the action with respect to the metric field only. In the spirit of gauge theories, one can also vary the action with respect to the connections and independently with respect to the metric. Then, it is possible to find second order corrections with no a priori background structures \cite{12}.

In some sense, our approach is similar to Horndeski's theory which is the most general one leading to second order equations of motion for gravity described by a metric $g_{\mu\nu}$ coupled with a scalar field $\phi $ and its first two derivatives \cite{13}. This theory involves non linear higher order derivatives of the scalar field, like $\big( \Box \phi \big)^2 $, and yet leads to second order. Moreover, if all the matter fields are minimally coupled with the same metric $\widetilde{g}_{\mu \nu} \big( g_{\mu\nu} , \phi \big)$, one can expect the equivalence principle to hold \cite{14}, which is also a fundamental feature of GR  that one wants to keep. 

Briefly, the outline of the paper is the following. First we consider all the independent  scalar invariants built from the metric field and its derivatives, for example of the form $\big( \Box R \big)^2$, and see if some linear combinations of them, or, in the spirit of  \cite{15} and \cite{16}, if some roots of these combinations, could lead to second order differential equations for FLRW space-time and static spherically symmetric one. The basis of independent scalars that are needed have been presented in \cite{17}, but for specific backgrounds, we will show that this basis may be reduced. Furthermore,  we will start to exibit, order by order for FLRW, the existence of polynomial and non-polynomial gravity models that give second order equations and polynomial corrections to the Friedmann equation. Finally, we will investigate the  static spherically symmetric space-times.


\section{Order 6 FKWC-basis}

The basis of all independent invariant geometrical scalars involving $2n$ derivatives of the metric are separated into different classes, depending on how many covariant derivatives act on curvature tensors. For order 6 ($n=3$), the first class, that does not involve explicitly covariant derivatives, from $\mathcal{L}_1$ to $\mathcal{L}_8$, is denoted by $\mathcal{R}_{6,3}^0$ : these scalars are built with six derivatives of the metric and by the contraction of 3 curvature tensors. The two other classes, $\mathcal{R}_{\left\{ 2,0 \right\}}^0$ and $\mathcal{R}_{\left\{ 1,1 \right\}}^0$, contain scalars that involve respectively, a curvature tensor contracted with two covariant derivatives acting on another curvature tensor (from $\mathscr{L}_1$ to $\mathscr{L}_4$), and two covariant derivatives, each acting on one curvature tensor (from $\mathscr{L}_5$ to $\mathscr{L}_8$) : 
~\\~
~\\~
$\left\{
\begin{array}{l}
 \mathcal{L}_1=R^{\mu\nu\alpha\beta}R_{\alpha\beta\sigma\rho}R^{\sigma\rho}_{\;\,\;\,\,\mu\nu} \quad\;\;\;\; , \;\;\quad  \mathcal{L}_2=R^{\mu\nu}_{\;\,\;\,\,\alpha\beta}R^{\alpha\sigma}_{\;\,\;\,\,\nu\rho}R^{\beta\rho}_{\;\,\;\,\,\mu\sigma}

\\  \mathcal{L}_3=R^{\mu\nu\alpha\beta}R_{\alpha\beta\nu\sigma}R^{\sigma}_{\;\,\mu} \quad\;\;\;\; , \;\;\quad  \mathcal{L}_4=R R^{\mu\nu\alpha\beta} R_{\mu\nu\alpha\beta} =R T

\\  \mathcal{L}_5=R^{\mu\nu\alpha\beta}R_{\mu\alpha}R_{\nu\beta} \;\;\quad\;\; , \;\;\quad  \mathcal{L}_6=R^{\mu\nu}R_{\nu\alpha}R^{\alpha}_{\;\,\mu}

\\  \mathcal{L}_7=R R^{\mu\nu} R_{\mu\nu} = R S \;\,\,  , \,\;\quad  \mathcal{L}_8=R^3
\end{array}
\right.
$
\\~
\\
\\
$\left\{
\begin{array}{l}
\mathscr{L}_1=R\Box R
 \quad \quad \;\;\; \quad\;\; , \;\quad 
\mathscr{L}_2=R_{\mu\nu}  \Box R^{\mu\nu} 

\\ \mathscr{L}_3=R^{\mu\nu\alpha\beta} \nabla_\nu \nabla_\beta  R_{\mu\alpha} 
\; \; , \;\;\,\,\,\;
 \mathscr{L}_4=R^{\mu\nu}  \nabla_\mu \nabla_\nu R

\\ \mathscr{L}_5=\nabla_\sigma R_{\mu\nu}  \nabla^\sigma R^{\mu\nu} 
\quad \quad\,\; , \;\;\;\; 
 \mathscr{L}_6=\nabla_\sigma R_{\mu\nu}  \nabla^\nu R^{\mu\sigma} 

\\ \mathscr{L}_7=\nabla_\sigma  R_{\mu\nu\alpha\beta}    \nabla^\sigma R^{\mu\nu\alpha\beta}
 \;\;\;\;\; , \;\;\;\;
 \mathscr{L}_8=\nabla_\sigma R \nabla^\sigma R
\end{array}
\right.
$
~\\~

Recall that our aim is to see that if we consider all these scalars, there are quite natural modified gravity Lagrangian densities that we can expect to lead to second order equations of motion and that actually do. There are linear combinations of all the scalars of the basis, but also for example square-root of the $\mathcal{R}_{\left\{ 1,1 \right\}}^0$-class, or cubic-roots of the $\mathcal{R}_{6,3}^0$ one, even if we are not going to study this last because we search here for high energy geometrical correction to the Einstein-Hilbert action. We write down this fact as :
\begin{eqnarray*}
 \mathscr{L}= \sum \Big( R^3 + R \nabla \nabla R +  \nabla R\nabla R \Big) + \sqrt{ \sum \big( \nabla R\nabla R \big) } + \sqrt[3]{\sum \big( R^3 \big) }.
\end{eqnarray*} 
Because inside a same class, the scalars have approximatively the same terms in their expansions, we can indeed expect to cancel higher order derivatives for some specific combinations of them, and then to have second order equations of motion.

Now, let us write down some definitions that allow to find relations between these scalars, coming from the fact that we are going to restrict our study to specific backgrounds, the Friedmann-Lema\^{i}tre-Robertson-Walker (FLRW) metric and the static spherically symmetric one, both in four dimensions. 

For both of them, there are between the scalars  relations coming from the Lovelock theorem (that are not taken into account in Ref  \cite{17}), and also for FLRW,  relations coming from the fact that this is a conformally invariant flat metric. All these relations are written in the first Appendix, and we need to express them to define the Weyl tensor, as : 
\begin{eqnarray}
W_{\mu\nu\alpha\beta}=R_{\mu\nu\alpha\beta}-\frac{1}{2}(R_{\mu\alpha}g_{\nu\beta}-R_{\mu\beta}g_{\nu\alpha}+R_{\nu\beta}g_{\mu\alpha}-R_{\nu\alpha}g_{\mu\beta})+\frac{1}{6}(g_{\mu\alpha}g_{\nu\beta}-g_{\mu\beta}g_{\nu\alpha})R.
\end{eqnarray}
And the following rank 2 tensor that is null in four dimension because of the Lovelock theorem :
\begin{eqnarray}
L_{\mu\nu} = -\frac{1}{2}g_{\mu \nu} \mathcal{E}_4 + 2 Q_{\mu \nu} -4 P_{\mu \nu} +4 R_{\ \nu \mu \ }^{\alpha \ \ \gamma} R_{\alpha \gamma} + 2 R R_{\mu \nu} =0,
\end{eqnarray}
where $Q_{\mu \nu} =R_{\mu\eta\alpha}^{\ \ \ \ \beta}R_{\nu \ \ \beta}^{\ \eta \alpha}$ and $P_{\mu \nu} =  R_{\nu\gamma}R^{\gamma}_{\ \mu}$.
Indeed, if we vary the Lagrangian associated with the Gauss-Bonnet invariant with respect to the metric field we find :
\begin{eqnarray}
\begin{split} 
\delta \big( \sqrt{-g}   \mathcal{E}_4 \big)  =&
\sqrt{-g} \; \delta g^{\mu \nu} L_{\mu\nu}.
\end{split}
\end{eqnarray}
But as we saw in equation (3), this Lagrangian can be written as a total derivative in four dimensions, which means that its contribution to the equations of motion is identically zero.

\section{Friedmann-Lema\^{i}tre Space-time} \vspace{-12pt}
\subsection{Order 6}

We start with the flat Friedmann-Lema\^{i}tre cosmological metric describing the dynamics of the universe at very large scale, in the simpliest manner : 
\begin{eqnarray}
ds^2=-dt^2 + a(t)^2 \big( dr^2 + r^2 d\Omega ^2 \big),
\end{eqnarray}
where $a(t)$ is the scale factor and $d\Omega^2 = d\theta ^2 + \sin ^2\theta \, d\phi ^2 $ is the metric of the 2-sphere. This metric is conformally invariant to Minkowski space-time, and  from the relations written in the first Appendix, we can choose the following reduced basis of all independent order 6  scalar invariants: $(\mathcal{L}_4, \mathcal{L}_6, \mathcal{L}_7,\curv{L}_1,\curv{L}_3,\curv{L}_5,\curv{L}_8)$. We note that there is one scalar less than for a general conformally invariant space-time coming from the particular metric of FLRW for which there is the additional relation: 
\begin{eqnarray}
\mathcal{L}_1=\frac{1}{3}\big( -\mathcal{L}_7+2 \mathcal{L}_4 \big).
\end{eqnarray}

\subsubsection{Linear Combination. $H^6$ correction.}

With this metric, we can right the most general order 6 linear combination of all the independent scalars : 
\begin{eqnarray*}
\begin{split} 
J=&\sum \Big( v_i \mathcal{L}_i +x_i \curv{L}_i \Big) =\frac{3}{a(t)^6} \Bigg( \sigma_1(v_i,x_i) \; \dot{a}(t)^6 + \sigma_2(v_i,x_i) \; a(t) \; \dot{a}(t)^4 \; \ddot{a}(t) + \sigma_3(v_i,x_i) \; a(t)^2 \; \dot{a}(t)^2 \; \ddot{a}(t)^2
   \\& + \sigma_4(v_i,x_i) \; a(t)^3 \; \ddot{a}(t)^3 + \sigma_5(v_i,x_i) \; a(t)^2 \; \dot{a}(t)^3 \; a^{(3)}(t)+ \sigma_6(v_i,x_i) \; a(t)^3 \; \dot{a}(t) \; \ddot{a}(t) \; a^{(3)}(t)
   \\& + \sigma_7(v_i,x_i) \; a(t)^4 \; a^{(3)}(t)^2+ \sigma_8(v_i,x_i) \; a(t)^3 \; \dot{a}(t)^2 \; a^{(4)}(t)+ \sigma_9(v_i,x_i) \; a(t)^4 \; \ddot{a}(t) \; a^{(4)}(t) \Bigg),
\end{split}
\end{eqnarray*}  
where the expressions of the $\sigma_j$ in terms of $(v_i,x_i)$ are presented in the second Appendix. 
Setting all of them to zero allows to check that our list of scalars is a basis. We can then impose $v_1=v_2=v_3=v_5=v_8=x_2=x_4=x_6=x_7=0$ to take into account the algebraic relations we have found. 

Moreover, in this section, we are interested in  linear combinations of order 6 scalars that lead to second order equations of motion. Therefore, we can also consider equivalence relations (up to boundary terms) between the scalars, and there are three of them that remain after considering the previous algebraic relations : 
\begin{eqnarray*}
 \int d^4x \sqrt{-g}  \curv{L}_3  = -\frac{1}{12} \int d^4x \sqrt{-g} \;  \curv{L}_8  \quad \quad \text{;} \quad  \int d^4x \sqrt{-g}  \curv{L}_1  = - \int d^4x \sqrt{-g} \; 
 \curv{L}_8 
\end{eqnarray*}
\begin{eqnarray*}
\text{and} \quad \int d^4x \sqrt{-g} \curv{L}_5   =  \frac{1}{6 }\int d^4x \sqrt{-g} \; \big( 2 \curv{L}_8 + 3 \mathcal{L}_4 - 12  \mathcal{L}_6 + \mathcal{L}_7 \big).
\end{eqnarray*}

We check that there is no other one by deriving the equations of motion for the scale factor considering the lagrangian  $L=a(t)^3 \; J$ , that we substitute into the generalized Euler-Lagrange equation for third order lagrangians: 
\begin{eqnarray}
-\frac{d^3}{dt^3} \Big( \frac{\partial L}{ \partial a^{(3)}} \Big) + \frac{d^2}{dt^2} \Big( \frac{\partial L}{ \partial \ddot{a}} \Big) - \frac{d}{dt} \Big( \frac{\partial L}{ \partial \dot{a}} \Big) + \frac{\partial L}{ \partial a} =0.
\end{eqnarray}

Finally we only need to consider the combination : 
\begin{eqnarray*}
J=v_4 \mathcal{L}_4 +v_6 \mathcal{L}_6 +v_7 \mathcal{L}_7 + x_8 \curv{L}_8  , 
\end{eqnarray*}
and after deriving the equation of motion and imposing a simultaneous cancellation of the higher order terms, we find that there is only one linear combination leading to second order equations :  
\begin{eqnarray}
J_1 =-7 \mathcal{L}_4 + 2 (6 \mathcal{L}_6 +\mathcal{L}_7)=72 H(t)^4 \left(3 \dot{H}(t)+2 H(t)^2\right),
\end{eqnarray}
where $H(t)=\dot{a}(t) / a(t)$ is the Hubble parameter. We note that this linear combination only involves contractions of curvature tensors, which is the kind of corrections expected to follow from quantum field theory.
Therefore, considering the following action that could represent a natural high energy geometrical correction to the Einstein-Hilbert action for FLRW space-time,  
\begin{eqnarray}
S_1= \int d^4x \sqrt{-g} \; \Bigg( \frac{1}{16 \pi}\bigg[ R +\nu  \Big(  -7 \, R R^{\mu\nu\alpha\beta}R_{\mu\nu\alpha\beta}  + 12\,  R^{\mu\nu}R_{\nu\alpha}R^{\alpha}_{\;\,\mu} + 2 \, R R^{\mu\nu} R_{\mu\nu}   \;\Big) \;\bigg] + \curv{L}_m \Bigg),
\end{eqnarray}
where $\curv{L}_m$ is the Lagrangian density for matter, one find the acceleration equation : 
\begin{eqnarray}
3 H(t)^2+2 \dot{H}(t)-36\, \nu \,  H(t)^4 \big(H(t)^2+2 \dot{H}(t)\big) = -8\pi p ,
   \end{eqnarray}
   where $p$ is the cosmic pressure. Then the equation of the conservation of the energy, with $\rho$ the cosmic energy density,
   \begin{eqnarray}
\frac{d\rho}{dt}+3H(t) \big(\rho + p \big)=0,
\end{eqnarray}
gives the following modified Friedmann equation : 
\begin{eqnarray}
3 H(t)^2-36 \, \nu \,   H(t)^6= 8 \pi \rho .
\end{eqnarray}
One can solve it, and choose only the solutions that reduce to the standard equation when $\rho$ is small. However, the only solution we are going to see in this work for FLRW spacetime is coming from a non-polynomial correction involving order 8 scalars, but the fact that $S_1$ is unique and second order could be a sufficient reason to study its cosmological solutions. 

\subsubsection{Non-polynomial gravity. $H^3$ correction.}

Now we want to consider all the order 6 linear combinations that are perfect squares, in order to consider non-polynomial corrections that are the square-roots of these squares. Rewrite the most general linear combination in terms of the Hubble parameter $H(t)$ : 
\begin{eqnarray*}
\begin{split} 
J=&\sum\limits_{i=4, 6, 7} v_i \mathcal{L}_i + \sum\limits_{j=1, 3, 5, 8} x_j \curv{L}_j  =3 \Bigg( \widetilde{\sigma}_1(v_i,x_i) \; H(t)^6 + \widetilde{\sigma}_2(v_i,x_i) \; H(t)^4 \dot{H}(t) + \widetilde{\sigma}_3(v_i,x_i) \; H(t)^2 \dot{H}(t)^2
   \\&  \quad \quad \quad \quad \quad \quad \quad \quad \quad \quad + \widetilde{\sigma}_4(v_i,x_i) \; \dot{H}(t)^3 + \widetilde{\sigma}_5(v_i,x_i) \; H(t)^3 \ddot{H}(t)+ \widetilde{\sigma}_6(v_i,x_i) \; H(t) \dot{H}(t) \ddot{H}(t)
   \\& \quad \quad \quad \quad \quad \quad \quad \quad \quad \quad + \widetilde{\sigma}_7(v_i,x_i) \ddot{H}(t)^2+ \widetilde{\sigma}_8(v_i,x_i) \; H(t)^2 H^{(3)}(t)+ \widetilde{\sigma}_9(v_i,x_i) \; \dot{H}(t) H^{(3)}(t) \Bigg).
\end{split}
\end{eqnarray*}  

To find squares, we use the following general procedure that is usefull for FLRW space-time, and necessary for spherical symmetry : 
Take the higher order perfect square $\ddot{H}(t)^2$. In the expansion of our square, each term will be multiply by $\ddot{H}(t)$, so the only terms that can enter inside are those $K_i(t)$ for which $K_i(t) \ddot{H}(t)$ and $K_i(t)K_j(t)$ exist in the expansion of order 6 scalars. Because of this, we need to impose the conditions : $  \widetilde{\sigma}_9\, =\, 0 \;$,  $\widetilde{\sigma}_8\, =\, 0 \, \;$ and $\, \widetilde{\sigma}_4\, =\, 0$, what give $x_1=x_3=0$ and $v_7=-v_4-5 v_6/12$. Therefore, $\widetilde{\sigma}_5\, =\, 0$ and there are only two possible forms of squares made of order 6 scalars : 
\begin{eqnarray*}
\sum\limits_{i,j}  \big( v_i \mathcal{L}_i + x_j \curv{L}_j   \big) = \Big( \delta H(t) \dot{H}(t) + \gamma H(t)^3 \Big)^2,
\end{eqnarray*}
\begin{eqnarray*}
\text{And :} \quad \quad \quad \sum\limits_{i,j}  \big( v_i \mathcal{L}_i + x_j \curv{L}_j   \big) = \Big( \xi  \ddot{H}(t) + \delta H(t) \dot{H}(t) \Big)^2.
\end{eqnarray*}

It means that all their square-roots can be decomposed in the basis $\big( H(t)^3, \, H(t) \dot{H}(t) , \, \ddot{H}(t) \big)$. Moreover, the general Lagrangian density :
\begin{eqnarray}
\sqrt{\sum\limits_{i,j}  \big( v_i \mathcal{L}_i + x_j \curv{L}_j   \big) }=  \xi  \ddot{H}(t) + \delta H(t) \dot{H}(t) + \gamma H(t)^3 \, ,
\end{eqnarray}
leads to second order differential equations for all $\big( \xi, \delta, \gamma \big) $, so we do not need to impose additional conditions on these coefficients and there are then only 3 independent second order corrections that we can find in this way.

Now let us see what are the actual perfect squares that one can find. Solving the natural conditions for respectively the first and the second kind of perfect squares, $\, \widetilde{\sigma}_2^2\, =\, 4 \, \widetilde{\sigma}_3\,  \widetilde{\sigma}_1 \, \;$, $\widetilde{\sigma}_6 = \widetilde{\sigma}_7 = 0$ and $\, \widetilde{\sigma}_6^2\, =\, 4 \, \widetilde{\sigma}_3\,  \widetilde{\sigma}_7 \, \;$, $\widetilde{\sigma}_1 = \widetilde{\sigma}_2 = 0$, one find the following ones : 
\begin{eqnarray}
\begin{split} 
&J_2 = 2 \mathcal{L}_4+12 \mathcal{L}_6 -7 \mathcal{L}_7 = -72 \bigg( 4 H^3 + 3 \dot{H}H \bigg)^2  \, ,
\\
& J_3= \bigg(6\mathcal{L}_4 -12\mathcal{L}_6-\mathcal{L}_7 \bigg)\alpha \big( 5\alpha +18\beta \big) +6 \big( \alpha +3 \beta \big) \bigg( \alpha  \curv{L}_5 + \beta \curv{L}_8 \bigg)
\\ &~~~~~ = -72 \bigg( 3 \big( \alpha +4\beta \big) H(t)\dot{H}(t) + \big( \alpha +3\beta \big) \ddot{H}(t) \bigg)^2 \, .
\end{split}
\end{eqnarray}
The last one being a general formula that gives perfect squares for any value of  $\big(\alpha,\beta\big)$.
As we just saw, there are only 3 independent square-roots of these squares, so from the general expression $J_3$, we can choose the two last to be the squares given by $\big(\alpha=0,\beta= \sqrt{2}/6 \big)$ and $\big(\alpha=1,\beta=0\big)$: 
\begin{eqnarray}
J_{3,1} =  \curv{L}_8  = -36 \left(4 H(t) \dot{H}(t)+\ddot{H}(t)\right)^2,
\end{eqnarray}
\begin{eqnarray}
\text{And :} \quad \quad \quad J_{3,2} = \Big( 6 \curv{L}_5+5 \big( 6 \mathcal{L}_4 - 12\mathcal{L}_6 -\mathcal{L}_7\big) \Big) =-72 \left(3 H(t) \dot{H}(t)+\ddot{H}(t)\right)^2.
\end{eqnarray}
Indeed, one can check that the following combination, with $\epsilon_i$ the signs inside de squares, 
\begin{eqnarray*}
\begin{split} 
\nu_0 \, \epsilon_0 \sqrt{-J_3} +\nu_1 \, \epsilon_1 \sqrt{-J_{3,1}}+\nu_2 \, \epsilon_2 \sqrt{-J_{3,2}} =& 6 \Big(3 \sqrt{2} \,  (\alpha +4 \,  \beta ) \,  \nu_0 \,  \epsilon_0+4 \,  \nu_1 \, 
   \epsilon_1+3 \,  \sqrt{2} \,  \nu _2 \,  \epsilon _2 \Big) H(t) \dot{H}(t) \\+& 6
   \Big( \sqrt{2} \,  (\alpha +3\,  \beta ) \, \nu _0 \, \epsilon _0+ \, \nu _1\,  \epsilon _1 \,+\sqrt{2} \, \nu _2 \, \epsilon _2 \Big) \ddot{H}(t) ,
\end{split} 
\end{eqnarray*}	
Vanishes for $\nu _2=-\alpha  \nu _0 \epsilon _0/\epsilon _2$ and $\nu _1=-3 \sqrt{2} \beta  \nu _0 \epsilon _0/\epsilon _1$, and so the relation becomes explicitly, for all value of $(\alpha,\beta)$ : 
	\begin{eqnarray*}
\begin{split} 
~& \sqrt{- \Bigg[ \Big(6\mathcal{L}_4 -12\mathcal{L}_6-\mathcal{L}_7 \Big)\alpha \big( 5\alpha +18\beta \big) +6 \big( \alpha +3 \beta \big) \Big( \alpha  \curv{L}_5 + \beta \curv{L}_8 \Big) \Bigg] }  \\ &- \frac{3\beta \epsilon_0 \sqrt{2}}{\epsilon_1} \sqrt{-\curv{L}_8 }  - 
 \frac{\epsilon_0 \alpha}{\epsilon_2}
 \sqrt{- \Big( 6 \curv{L}_5+5 \big( 6 \mathcal{L}_4 - 12\mathcal{L}_6 -\mathcal{L}_7\big) \Big) } =0.
\end{split} 
\end{eqnarray*}
Therefore, this general formula for perfect squares depends only on $\sqrt{-J_{3,1}}$ and $\sqrt{-J_{3,2}}$ as we said. 
\\ We note here that an interesting property coming from the existence of an infinite number $J_3\big(\alpha,\beta\big)$ of perfect squares for which the square-root can be decomposed in a small basis is that it gives some non-linear algebraic relations between the scalars of the FKWC-Basis, that reduce it in a non-trivial way, and allow to have a very small number of independent corrections. Indeed, solving the previous equation for $\mathcal{L}_4$, we find :
\begin{eqnarray}
\begin{split} 
\mathcal{L}_4 = \frac{1}{54} \Big( -9  \curv{L}_5 + 2  \curv{L}_8 + 108 \mathcal{L}_6 + 9 \mathcal{L}_7 - \sqrt{ \curv{L}_8 \big( 18  \curv{L}_5 -5 \curv{L}_8 \big) } \epsilon_0 \epsilon_1 \Big).
\end{split} 
\end{eqnarray}
We can now calculate the equations of motion for the 3 Lagrangian densities $\sqrt{-J_{2}}$, $\sqrt{-J_{3,1}}$ and $\sqrt{-J_{3,2}}$. First, one can check that the last one is in fact a topological term that does not bring any contribution to the equation of motion. Moreover, the first two lagrangian densities give the same equation of motion $54 \, \dot{a}(t) \ddot{a}(t)=0$ for the first one, and $18 \sqrt{2} \, \dot{a}(t) \ddot{a}(t)=0$ for the second one. It means that they are equal up to an invariant scalar $T$ for which $\sqrt{-g T}$ is a total derivative, 
 \begin{eqnarray}
\sqrt{- \Big(2 \mathcal{L}_4+12 \mathcal{L}_6 -7 \mathcal{L}_7\Big) }=\frac{3}{\sqrt{2}} \sqrt{- \curv{L}_8} + T \; \, \text{,}
\end{eqnarray}
 such that we can in fact consider a unique scalar (let us choose $\sqrt{- \curv{L}_8}$) made of order 6 scalars that leads to non vanishing second order differential equations. We can note that $T$ cannot be equal to $\sqrt{-J_{3,2}}$ because $\sqrt{-J_{2}}$ contains an $H^3$ terms.
 Therefore $T$ could be found considering higher order derivatives scalars, and other perfect powers. In this case, powers of four for order 12 scalars for example. There is then another non-linear relation between order 6 scalars and (possibly) order 12 ones. 

Finally, to recapitulate this part, we have found that it is natural to consider only one perfect square made of order 6 scalars : $\curv{L}_8 =  \nabla^\sigma R \nabla_\sigma R $. As a result,  the action :
\begin{eqnarray}
S_2= \int d^4x \sqrt{-g} \; \Bigg( \frac{1}{16 \pi}\bigg[ R +\nu  \sqrt{ - \nabla^\sigma R \nabla_\sigma R }   \;\bigg] + \curv{L}_m \Bigg),
\end{eqnarray}
leads to a unique second order $H^3$-correction to the Friedmann equation, as it is easy to see using the same reasoning as in the previous section. 

\subsection{Order 8}

Now let us study the linear combination and squares made of order 8 scalars that lead to second order equations of motion. We do not copy all the FKWC basis for general metric, but we name the scalars according to their position in Ref \cite{17} where this basis is fully written. The reduced FKWC basis for order 8 scalars in FLRW space-time is the following : 
~\\~

$\left\{
\begin{array}{l}
\mathcal{K}_1=R^4
\quad\; , \quad  
\mathcal{K}_{10}=R^{\mu\nu}R^{\alpha\beta}R^{\sigma\rho}_{\;\,\;\,\,\mu\alpha}R_{\sigma\rho\nu\beta}
\quad\; , \quad  
\mathcal{K}_{11}=R\,R^{\mu\nu\alpha\beta}R_{\mu\;\,\alpha}^{\;\,\sigma\;\,\rho}R_{\nu\sigma\beta\rho} = R \big( \frac{1}{4} \mathcal{L}_1 -\mathcal{L}_2 \big)
\\~
\\
 \mathcal{K}_{12}=T^2 
 \quad\; , \quad  
 \mathcal{M}_1 = R\,\Box^2R
 \quad\; , \quad  
 \mathcal{M}_2 =R_{\mu\nu} \nabla^{\mu}\nabla^{\nu}\Box R 
 \quad\; , \quad  
 \mathcal{M}_3=R^{\mu\nu}\Box^2R_{\mu\nu}
 \\~
\\
 \mathcal{M}_5=\nabla^{\mu}\Box R \nabla_{\mu} R
 \;\;\, , \;\;  
 \mathcal{M}_6=\nabla_{\mu}\nabla_{\nu}\nabla_{\alpha}R \nabla^{\mu} R^{\nu\alpha}
  \;\;\, , \;\;  
   \mathcal{M}_{10}=\big(\Box R \big)^2
   \;\; \,  , \;\;  
    \mathcal{M}_{11}=\nabla_{\mu}\nabla_{\nu}R \nabla^{\mu}\nabla^{\nu}R 
 \\~
\\
 \mathcal{M}_{12}=\nabla^{\mu}\nabla^{\nu}R  \Box R_{\mu\nu}
 \quad\; , \quad  
 \mathcal{M}_{14}=\nabla_{\mu}\nabla_{\nu} R_{\alpha\beta} \nabla^{\mu}\nabla^{\nu} R^{\alpha\beta} 
  \quad\; , \quad 
  \mathcal{M}_{18}=R \, \curv{L}_1
   \quad\; , \quad 
   \mathcal{M}_{19}=R \, \curv{L}_4
   \\~
\\ \mathcal{M}_{20}= S \Box R 
    \quad\; , \quad 
\mathcal{M}_{33}= R  \, \curv{L}_8
\end{array}
\right.
$
\\~
\\~~~~~~\\

We also introduce the definitions 
$\mathcal{K}_{9}=R^{\mu\nu}R^{\alpha}_{\;\,\mu}R_{\nu}^{\;\,\beta\sigma\rho} R_{\rho\sigma\beta\alpha}$ $\,$, $\,$ $\mathcal{M}_{13}=\Box R_{\mu\nu} \Box R^{\mu\nu}$ $\,$ and $\,$ $\mathcal{M}_{16}=\nabla_{\mu}\nabla_{\nu} R_{\alpha\beta} \nabla^{\beta}\nabla^{\alpha} R^{\nu\mu}$
that will be usefull later for static spherically symmetric space-times.

\subsubsection{Linear Combination. $H^8$ correction.}
Consider the sum of all independent order 8 scalars for FLRW space-time : 
\begin{eqnarray*}
J=&\sum  v_i \mathcal{K}_i + \sum x_j \mathcal{M}_j 
\end{eqnarray*}  
Here, we follow exactly what we did for order 6 scalars. We derive the equation of motion associated with the previous sum, and see what conditions on $(v_i , x_j)$ cancel the equation, such that we find the 10 equivalence relations that exist between the scalars of the reduced basis. Therefore, we can consider only the following independent scalars with respect to the equation of motion, $\big( \mathcal{K}_{1},\mathcal{K}_{10},\mathcal{K}_{11},\mathcal{K}_{12},\mathcal{M}_{1},\mathcal{M}_{11},\mathcal{M}_{12} \big)$ and the reduced sum : 
\begin{eqnarray*}
J=&\sum\limits_{i=1, 10, 11, 12}   v_i \mathcal{K}_i + \sum\limits_{j=1, 11, 12}  x_j \mathcal{M}_j  \, .
\end{eqnarray*}  
We derive its associated equation of motion and see what a simultaneous cancellation of all the higher order terms implies for the coefficients $(v_i , x_j)$ : we find that the unique linear combination of order 8 scalars for FLRW space-time that leads to second order equation is : 
\begin{eqnarray}
J_{4}=\mathcal{K}_{1} -48 \, \mathcal{K}_{11} -9 \, \mathcal{K}_{12} =1728 \, \Big( \, H^8 +2 \dot{H}H^6 \,  \Big).
\end{eqnarray}
Therefore, one may consider the action : 
\begin{eqnarray}
\begin{split} 
S_{3} = \int d^4x \sqrt{-g} \; \Bigg( \frac{1}{16 \pi}\bigg[ R +\nu \Big( R^4 -48 \, R\,R^{\mu\nu\alpha\beta}R_{\mu\;\,\alpha}^{\;\,\sigma\;\,\rho}R_{\nu\sigma\beta\rho} -9 \, \big( R^{\mu\nu\alpha\beta} R_{\mu\nu\alpha\beta} \big)^2 \Big)   \;\bigg] + \curv{L}_m \Bigg) ,
\end{split}
\end{eqnarray}
and see that it brings an $H^8$ correction to the Friedmann equation. We note that this correction involves only contraction of curvature tensors like for the order 6 case.

\subsubsection{Non-polynomial gravity. $H^4$ correction.}

\paragraph{Correction to the Einstein-Hilbert action}

~\\~
\indent To find non-polynomial second order models from order 8 scalars, we follow exactly what we did for the order 6 and find the same kind of result : there are only two classes of perfect squares in this case. Those for which the square-roots give topological scalars, that does not give any contribution to the equation of motion, and a class of equivalent scalars with respect to the equation of motion, up to the topological scalars of the first class. Therefore in this case also, we can consider a unique perfect square that contributes to the dynamics.

To begin, the more general square made of order 8 scalars has the form :
\begin{eqnarray*}
\big( \alpha  H(t)^4+\beta  H(t)^2 \dot{H}(t)+\gamma  \dot{H}(t)^2+\delta  H(t)
   \ddot{H}(t)+\sigma  H^{(3)}(t) \big)^2.
\end{eqnarray*}  
And its square-root gives the following equation of motion: 
   \begin{eqnarray*}
   \begin{split}
~& 3 \big( \alpha -\beta +\gamma +2 \delta -6 \sigma \big) \, \dot{a}(t)^2 \, \Big( \dot{a}(t)^2-4
   a(t) \ddot{a}(t)\Big)
   \\
 &+\big( \gamma -\delta +3 \sigma \big)  \, a(t)^2 \, \Big(3
   \ddot{a}(t)^2+4 \dot{a}(t) a^{(3)}(t)+2 a(t) a^{(4)}(t)\Big) =0.
   \end{split}
\end{eqnarray*}
Following the section concerning order 6, we can say that there are 5 independent contributions made of square-roots of order 8 scalars, but in this case, there is also one condition to impose on  $\big( \alpha, \beta, \gamma, \delta, \sigma \big)$ in order to have second order equations of motion. Therefore there will be only 4 independent contributions and we can choose them to be : 
  \begin{eqnarray}
   \begin{split}
&    \sqrt{J_5} = \sqrt{ -38 \mathcal{K}_1-2448 \mathcal{K}_{10}+2400 \mathcal{K}_{11}+1086 \mathcal{K}_{12}-143
   \mathcal{M}_{10}-220 \mathcal{M}_{11}+792 \mathcal{M}_{12}+88 \mathcal{M}_{18}-352 \mathcal{M}_{19} } 
   \\ &~~~~~~~=6 \sqrt{33} \left(3 H(t) \ddot{H}(t)+H^{(3)}(t)\right) ,
   \\& 
\mathcal{E}_4 = \sqrt{ \big( \mathcal{E}_4 \big)^2} =\frac{1}{\sqrt{33}} \sqrt{\mathcal{K}_{1} -144\mathcal{K}_{10}+48 \mathcal{K}_{11} +27 \mathcal{K}_{12}}  =24 \,H(t)^2 \left(H(t)^2+\dot{H}(t)\right) ,
\\& 
\Box R = \sqrt{\mathcal{M}_{10} } = -6 \left(12 H(t)^2 \dot{H}(t)+4 \dot{H}(t)^2+7 H(t)
   \ddot{H}(t)+H^{(3)}(t)\right) ,
   \\& 
  \sqrt{  J_6 }=\sqrt{ - \Big( 5\mathcal{K}_{1} + 9 \big( 8 \mathcal{K}_{10}-32 \mathcal{K}_{11} -7 \mathcal{K}_{12} \big) \Big) }= 12 \sqrt{66} H(t)^2 \dot{H}(t) ,
\end{split}
\end{eqnarray}
where only the last one gives a non vanishing contribution to the equations of motion. We note that, of course, because there are much more perfect squares than the previous ones, the fact that they form a basis gives once more some non-linear algebraic relations between the order 8 scalars for FLRW space-time, and reduces the already reduced FKWC-basis. However it is not the aim of this work to find all these relations. We focus here on the unique modified action with respect to order 8 scalars : 
\begin{align}
S_4=& \int d^4x \sqrt{-g} \; \Bigg( \frac{1}{16 \pi}\bigg[ R +\nu 
\sqrt{ 
-5 R^4 -9 \Big( 8 R^{\mu\nu}R^{\alpha\beta}R^{\sigma\rho}_{\;\,\;\,\,\mu\alpha}R_{\sigma\rho\nu\beta} - 32 R\,R^{\mu\nu\alpha\beta}R_{\mu\;\,\alpha}^{\;\,\sigma\;\,\rho}R_{\nu\sigma\beta\rho}  
}  \nonumber\\
 & \quad\quad\quad\quad\quad\quad\quad\quad\quad\quad\quad \overline{ ~ -7 \big( R^{\mu\nu\alpha\beta} R_{\mu\nu\alpha\beta} \big)^2  ~\Big) 
 }   \;\bigg] + \curv{L}_m \Bigg). \nonumber\\
\end{align}
\paragraph{Correction to the Friedmann equation}
~\\~
\indent From this action, we get the acceleration equation : 
\begin{eqnarray}
3 H(t)^2+2 \dot{H}(t)-6 \nu \sqrt{66} H(t)^2 \left(3 H(t)^2+4 \dot{H}(t)\right) = -8\pi p .
   \end{eqnarray}
In analogy with equations (12) to (14), this last modification leads to an $H^4$ modification of the Friedmann equation. Now we are going to see that this unique correction is very interesting regarding the problem of the Big Bang singularity as it was shown in Ref \cite{16}.   
It is convenient to introduce the dimensional quantity $\rho_c$,  critical energy density of the universe, and the constant parameter $ \epsilon $, such that
\begin{eqnarray}
\frac{\epsilon}{8 \pi \rho_c}=- 2 \nu \sqrt{66}\,.
\end{eqnarray}
Thus,  the modified Friedmann equation induced by the action $S_4$ is :
\begin{eqnarray}
3 H(t)^2+ \epsilon \; \frac{9  H(t)^4}{8 \pi \rho_c}=8 \pi 
   \rho.
\end{eqnarray}
The solution  $H(t)^2$  that reduces to the standard Friedmann equation in the limit when $ \rho_c $ goes to infinity  is : 
\begin{eqnarray}
3 H(t)^2 = \, \frac{4 \pi  \rho_c }{ \epsilon }\left(-1+\sqrt{\frac{4 \epsilon  \rho }{\rho_c}+1} \, \right).
\end{eqnarray}
Furthermore, if the energy density of the universe is small compared to the critical one, $\rho \ll \rho_c$, this equation becomes : 
\begin{eqnarray}
H(t)^2 = \, \frac{8 \pi \rho }{3} \Big( 1 - \epsilon \, \frac{\rho}{\rho_c} \Big) + O(\rho^3).
\end{eqnarray}
Choosing $\epsilon=1$, gives the loop quantum cosmology correction to the Friedmann equation  \cite{19}, and choosing $\epsilon=-\frac{1}{2}$, gives the Randall-Sundrum brane world model \cite{20}. These two corrections are regular cosmological solutions and allow to avoid the Big-Bang singularity \cite{19,16}. One should also note that the above equation may be obtained within other approches (see \cite{21,22}).

\section{Static Spherically Symmetric Space-times}

In this section,  we study  static spherically symmetric space-times, defined by the general metric:
\begin{eqnarray}
ds^2 = -B(r) dt^2 + A(r) dr^2 + r^2 d\Omega^2.
\end{eqnarray}
First, following exactly the same procedure as for the FLRW case, it is possible to show that there is no second order linear combination made of order 6 scalars. The same conclusion is valid for the classes $\mathcal{R}_{2,2}^0$ and $\mathcal{R}_{8,4}^0$. 

Now for order 4 scalars (and more generally for all orders, considering only monomials of the curvature tensor), the result of Deser and Ryzhov \cite{18} shows that the most general second order action is :
\begin{eqnarray}
S_5 = \int d^4x \sqrt{-g} \;  \Big(  R +\, \sqrt{3} \,  \sigma \,  \sqrt{ W^{\mu\nu\alpha\beta}W_{\mu\nu\alpha\beta}}  \; \Big).
\end{eqnarray}
Moreover, there is no other perfect squares than $W^{\mu\nu\alpha\beta}W_{\mu\nu\alpha\beta}$ and $R^2$. Concerning order 4, all the scalars that are perfect squares lead to second order equations of motion inside the square.

Now, starting from order 6 scalars, it is possible to show that there are again only two perfect squares. Thus,   we can consider the action : 
\begin{eqnarray}
S_6 = \int d^4x \sqrt{-g} \;  \Bigg(  \delta \,   \sqrt{ \nabla_\sigma R \nabla^\sigma R } + \, \sqrt{3} \, \gamma \,  \sqrt{ C^{\mu\nu\alpha}C_{\mu\nu\alpha}}  \; \Bigg),
\end{eqnarray}
here $C_{\mu\nu \alpha}$ is the Cotton tensor, expressed in terms of the Weyl tensor as $C_{\mu\nu \alpha}  = -2 \nabla^{\sigma} W_{\mu \nu\alpha\sigma} $. Its square may be  written in our basis as $C^{\mu\nu\alpha}C_{\mu\nu\alpha}=2 \big( \curv{L}_5-\curv{L}_6 \big) - \frac{1}{6} \curv{L}_8$.

In our search for second order differential equations, it is interesting to note that the property of order 4 perfect squares is preserved here : these two terms are such that their higher order terms cancel perfectly, what makes their associated equations of motion second order. For example the term : 
\begin{eqnarray}
\begin{split} 
\sqrt{-g} \;  &\sqrt{ C^{\mu\nu\alpha}C_{\mu\nu\alpha} } 
 =\frac{\sqrt{3}}{6} \frac{\sqrt{B(r)}}{r B(r)^3A(r)^3} \Bigg( \Sigma \Big(r,B(r),A(r),B'(r),A'(r),B''(r),A''(r) \Big)
   \\ &-3 r^3 B(r)^2 A(r) A'(r) B''(r)  
-r^3 B(r)^2 A(r) B'(r) A''(r)+2 r^3 B(r)^2 A(r)^2
   B^{(3)}(r) \Bigg),
\end{split}
\end{eqnarray}
where $\Sigma \Big(r,B(r),A(r),B'(r),A'(r),B''(r),A''(r) \Big)$ is a sum of 15 first order terms (that lead trivially to second order differential equations), is equivalent, up to boundary terms, to the following first order expression : 
\begin{eqnarray}
\begin{split} 
\sqrt{-g} \;  &\sqrt{ C^{\mu\nu\alpha}C_{\mu\nu\alpha} }  \equiv \frac{\sqrt{3}}{6} \frac{\sqrt{B(r)}}{r B(r)^3A(r)^3} \Bigg( 4B(r)^3 A(r)^2-4 B(r)^3 A(r)^3-5 r B(r)^2 A(r)^2 B'(r)
\\&\quad\quad \quad\quad\quad\quad\quad\quad\quad\quad\quad\quad\quad\quad\quad +2 r^2
   B(r) A(r)^2 B'(r)^2-\frac{1}{4} r^3 A(r)^2 B'(r)^3 \Bigg).
\end{split}
\end{eqnarray}  
Therefore, we have shown first that, up to order 6, all the perfect squares that one can build for static spherically symmetric space-time are also perfect squares in FLRW as we have seen for the case of $ \nabla_\sigma R \nabla^\sigma R$, because in this space-time  $W_{\mu \nu\alpha\sigma} =0$. And secondly, that they also share the property that their square-root lead to second order equations of motion for the metric field. 

These perfect squares, coming from the action $S_6$ for $A(r)$ and $B(r)$ are  respectively: 
\begin{eqnarray}
\begin{split} 
~&  16 \big(\gamma +2 \delta \big)
   B(r)^3+4 r \big(-5 \gamma +2 \delta \big) B(r)^2 B'(r)
   \\& +4 r^2 \big(2 \gamma
   +\delta \big) B(r) B'(r)^2+r^3 \big(-\gamma +\delta \big) B'(r)^3=0,
\end{split}
\end{eqnarray}  
And, 
\begin{eqnarray}
\begin{split} 
~& 8 \big(\gamma +2 \delta \big) \Big( -1+A(r)\Big) A(r) B(r)^3+4 r \big(5 \gamma -2 \delta
   \big) B(r)^3 A'(r)-18 r^2 \gamma  A(r) B(r) B'(r)^2
   \\& +8 r \big(2 \gamma
   +\delta \big) B(r)^2 \Big(-r A'(r) B'(r)+A(r) \left(B'(r)+r
   B''(r)\right)\Big)
   \\& +r^3 \big(\gamma -\delta \big) B'(r) \Big(5 A(r)
   B'(r)^2+3 B(r) \left(A'(r) B'(r)-2 A(r) B''(r)\right)\Big)=0.
\end{split}
\end{eqnarray}  
The first one provides solutions for $B(r)$, so the two equations decouple. 
We have found real exact vacuum solutions for 3 couples $(\gamma \, , \, \delta)$ and have computed their associated Kretschmann scalars $ R^{\mu\nu\alpha\beta}R_{\mu\nu\alpha\beta}$ to see if they suffer from singularites at $r=0$.
\begin{itemize}
   \item For $\gamma =1$ and $\delta =0$ :  
\begin{align} 
 ds^2 = -k \, r^2 \; dt^2 + \frac{r^2}{p+r^2} \; dr^2 + r^2 d\Omega^2 \quad\quad \text{and} \quad\quad R^{\mu\nu\alpha\beta}R_{\mu\nu\alpha\beta} \to  \frac{p^2}{r^8} ,
    \end{align}  
 \item For $\gamma =0$ and $\delta =1$ :  
\begin{align} 
 ds^2 = -\frac{k}{r^4} \; dt^2 + \frac{3}{1+p \, r^2} \; dr^2 + r^2 d\Omega^2 \quad\quad \text{and} \quad\quad R^{\mu\nu\alpha\beta}R_{\mu\nu\alpha\beta} \to  \frac{15 p^2-2 p+3}{r^4} ,
    \end{align} 
 \item For $\gamma =1$ and $\delta =-\frac{1}{2}$ :
\begin{align} 
 ds^2 = -k \; dt^2 + p \;dr^2 + r^2 d\Omega^2 \quad\quad \text{and} \quad\quad R^{\mu\nu\alpha\beta}R_{\mu\nu\alpha\beta} \to  \frac{(p-1)^2}{p^2 \, r^4} .
    \end{align}  
\end{itemize}
where $p$ and $k$ are integration constants. Note that $k$ can be set equal to one by a right choice of the time coordinate. 

Recall that the Kretschmann scalar of the Schwarzschild solution of the Einstein equations diverges as $ R^{\mu\nu\alpha\beta}R_{\mu\nu\alpha\beta} \to 1/r^6$. Therefore, we see that our first solution has a milder divergence. In the last case, choosing the initial condition $p=1$ provides $R^{\mu\nu\alpha\beta}R_{\mu\nu\alpha\beta}=0$ for all $r$ so there is no singularity of curvature because the space-time is flat. However in the second case, as for $p$ real, $15 p^2-2 p+3 >0$, we cannot "cancel" the singularity by a proper choice of initial condition, but still, the divergence is again milder than the standard one.

Now let us note that considering only the two order 8 classes $\mathcal{R}_{2,2}^0$ and $\mathcal{R}_{8,4}^0$, we have found a unique perfect square $\Big[ 2K_{10}-2K_{9}-M_{11}+2M_{12}-M_{13}+4M_{14}-4M_{16}\Big]$ (that is zero in the FLRW case, so possibly again related with Cotton and Weyl "conformal" tensors), but its square-root does not lead to second order equations of motion, such that we can conjecture that this property is true only for the 3 scalars $W^{\mu\nu\alpha\beta}W_{\mu\nu\alpha\beta}$, $C^{\mu\nu\alpha}C_{\mu\nu\alpha}$ and $ \nabla_\sigma R \nabla^\sigma R $, without counting the scalars $R^2$, $(\Box^i R)^2$, $(\mathcal{E}_4)^{2}$, etc, for which the square-root always lead to second order equations. 

To conclude this section, we point out that it would be interesting to find black hole solutions, for some values of $\sigma$, $\gamma$ and $\delta$, and starting from the following action :
\begin{eqnarray}
S_7 = \int d^4x \sqrt{-g} \;  \Big(  R + \, \sqrt{3} \, \sigma \,  \sqrt{ W^{\mu\nu\alpha\beta}W_{\mu\nu\alpha\beta}}  +\, \sqrt{3} \, \gamma \,  \sqrt{ C^{\mu\nu\alpha}C_{\mu\nu\alpha}} +  \delta \,   \sqrt{ \nabla_\sigma R \nabla^\sigma R } \; \Big),
\end{eqnarray}
where the two last terms would be consider as high energy corrections to the standard Einstein-Hilbert action. In this action, the two additional terms are very similar to one present in this $S_5$. We recall that starting from $S_5$,  one can find exact black hole solutions \cite{15}.

 After finding black hole solutions for some class of the parameters, it would be possible to reduce the number of physically relevant value of $\big( \sigma, \,\gamma, \,\delta \big)$, for example by calculating the Wald entropy associated with the action and the class of solutions, like it is done in Ref \cite{23}, by imposing the positivity of the entropy to find some contraint on the parameters. 

\section{Conclusions}

Considering the order six and eight FKWC-basis of all independent invariant scalars build from the metric field and its derivatives, we have found modified gravity models admitting second order equations of motion for FLRW and static spherically symmeric spacetimes.

For FLRW, we have shown that there are unique second order polynomial corrections to the Einstein-Hilbert action for both order six and eight. They involve only contractions of curvature tensors which is what one could expect to follow from quantum field theory. Yet, concerning static spherical symmetry we have seen that there is none.

Some non-polynomial corrections, that are square-roots of the specific combinations that are perfect squares have also been studied for FLRW and spherical symmetry for both order six and eight. We have shown that up to order 6, all the perfect squares lead to second order equations of motion in this way. However, for order eight, there are squares that do not have this property, for example the one we have found for spherical symmetry that involves scalars of the $\mathcal{R}_{2,2}^0$ and $\mathcal{R}_{8,4}^0$ order eight classes.

Moreover, concerning FLRW we have seen a "mechanism" that provides unique corrections with non-vanishing contributions to the equations of motion : there are lots of squares, but their square-roots can be decomposed in a very small basis, such that there exist some non-linear algebraic relations between order six and order eight scalars. Taken into account, we can choose the independent "square-root" scalars such that for both order, there are topological scalars and only one that is not.

For order six, the one we choose turned out to be the same as for spherical symmetry. The other perfect squares for this last spacetime are squares of the Weyl and Cotton tensors, which are identically zeros for FLRW, such that all the squares of static spherical symmetry are present in FLRW. Let's say also that the ressemblance between $S_6$ and $S_5$, that admits exact black hole solutions, could suggest that it might be possible to find also these kind of solutions from this first. It could also be interesting to understand better why the second order corrections specific to spherical symmetry involve "conformal" tensors.

We note here that in addition, we have checked if this common non-vanishing square that is $\nabla_\sigma R \nabla^\sigma R$ is also a square for another physically relevant spacetime, that is Bianchi I, defined by the following metric and describing an anisotropic universe as the early universe was \cite{28,29} : 
\begin{eqnarray}
ds^2= -dt^2 + \alpha(t) dx^2 + \beta(t) dy^2 + \gamma(t) dz^2.
\end{eqnarray}
It turns out that this scalar is also a perfect square in this case and that it leads also to second order equations of motions. These results are just copied in the third Appendix, and makes the results concerning the action :
\begin{eqnarray}
S= \int d^4x \sqrt{-g} \; \Bigg(  R +\nu  \sqrt{ - \nabla^\sigma R \nabla_\sigma R }   \Bigg),
\end{eqnarray}
our main ones.

The study of non-polynomial gravity with order eight scalars represents also an interesting results in the sense that we can choose the four independent square-root scalars to be three topological ones, and one that contributes to the equations of motion and gives the well-known $H^4$ correction to the Friedmann equation \cite{26,27}, that leads to a unique physically relevant solution for which the limit of small density allows to reproduce the loop quantum cosmology result and the one coming from the Randall-Sundrum brane world model. In these two, the big-bang singularity is absent and the cosmological solutions are regular ones.

To finish, we note that our study was done on fixed backgrounds, but General Relativity is a background independent theory, so it could be important to check for general background if the scalar $\nabla_\sigma R \nabla^\sigma R$ is still a perfect square, and if it is the case, if its square-root leads in the general case to second order equations of motion for the metric field. Moreover, to our knowledge, there is no proof that, considering all the FKWC-basis for all orders, it is impossible to find linear combinations of invariant scalars that lead to second order partial differential equations for a general metric. Indeed the Lovelock papers \cite{10,30,31} are only restricted their studies to lagrangians of the general form: $ L \Big(g_{\mu\nu} , \partial_{\alpha}g_{\mu\nu} ,\partial_{\alpha}\partial_{\beta}g_{\mu\nu}\Big)$, i.e. involving curvature tensors, but not explicit covariant derivatives of them. Therefore, if this remark is true, one could search for second order equations of motion coming from the general lagrangian $L \Big(g_{\mu\nu} , \partial_{\alpha}g_{\mu\nu} , ... , \partial_{\alpha}... \partial_{\beta} g_{\mu\nu} \Big)$. We know from our results that there is no such combination for order six and eight scalars because there is none for static spherically symmetric spacetime. But as the order grows, even if there are more and more terms in the expansions of the scalars, the number of independent scalars inside a same class grows as well, such that for some "high" order in the FKWC-basis, it might be possible to cancel all the higher order derivatives for general metric.



\pagebreak
\section*{Acknowledgments}







\appendix
\section{Appendix}
\vspace{-12pt}
\subsection{Relations for the reduced FKWC basis}

First we copy the following relations from Ref \cite{24} :
\begin{eqnarray}
\mathcal{E}_6=2 \mathcal{L}_1+8 \mathcal{L}_2+24 \mathcal{L}_3+2 \mathcal{L}_4+24 \mathcal{L}_5+16 \mathcal{L}_6-12 \mathcal{L}_7+ \mathcal{L}_8
\end{eqnarray}
is the order 6 Euler density,
\begin{eqnarray}
 W^2=W_{\mu\nu\alpha\beta} W^{\mu\nu\alpha\beta}=\frac{1}{3}R^2-2S+T
\end{eqnarray}
is the square of the Weyl tensor,
\begin{eqnarray}
W^3_1=W^{\mu\nu\alpha\beta}W_{\alpha\beta\sigma\rho}W^{\sigma\rho}_{\;\,\;\,\,\mu\nu}
\end{eqnarray} 
and
\begin{eqnarray}
W^3_2=W^{\mu\nu\alpha\beta}W_{\mu\sigma\rho\beta}W_{\;\;\;\nu\alpha}^{\sigma\;\,\;\,\,\;\rho}
\end{eqnarray} 
are the two independent cubic contractions of the Weyl tensor in four dimensions. 

From the relations of Ref \cite{25}, we have found the following geometrical identities :
\begin{eqnarray}
L_{\mu\nu}R^{\mu\nu}=-(2 \mathcal{L}_3 + \frac{1}{2}\mathcal{L}_4+4\mathcal{L}_5+4\mathcal{L}_6-4\mathcal{L}_7+\frac{1}{2}\mathcal{L}_8)
\end{eqnarray}
\begin{eqnarray}
 W_{\mu\nu\alpha\beta}\nabla^{\nu} \nabla^{\beta}R^{\mu\alpha}=\mathscr{L}_3-\frac{1}{2}\mathscr{L}_2+\frac{1}{12}\mathscr{L}_1+\frac{1}{2}\mathcal{L}_6-\frac{1}{2}\mathcal{L}_5
\end{eqnarray}
\begin{eqnarray}
\nabla^\rho R^{\mu\nu\alpha\beta}\nabla_{\rho}W_{\mu\nu\alpha\beta}=\mathscr{L}_7-2\mathscr{L}_5+\frac{1}{3}\mathscr{L}_8
\end{eqnarray}
\begin{eqnarray}
R^{\mu\alpha}\nabla^{\beta}\nabla^{\nu}W_{\mu\nu\alpha\beta}=\frac{1}{2}\curv{L}_2-\frac{1}{12}\curv{L}_1-\frac{1}{6}\curv{L}_4-\frac{1}{2}L_6+\frac{1}{2}L_5
\end{eqnarray}
\begin{eqnarray}
\nabla^\mu W_{\mu\nu\alpha\beta}\nabla^{\alpha} R^{\nu\beta}=\frac{1}{2}\curv{L}_5-\frac{1}{2}\curv{L}_6-\frac{1}{24}\curv{L}_8
\end{eqnarray}
Note that $\nabla^\mu \nabla^\nu L_{\mu\nu}=0$ identically, such that there are no new relations coming from this term. 

Because in four dimensions, $\mathcal{E}_6=0$ and $L_{\mu\nu}=0$, leading to $L_{\mu\nu}R^{\mu\nu}=0$ for order 6 scalars, the FKWC basis, that does not take into account these relations, is then reduced. 

As for conformally invariant space-times, their metrics verify $W_{\mu\nu\alpha\beta}=0$ which provides again new relations between the scalars coming from :

 $R \, W^2 =0$, $W^3_1 =0$, $W^3_2 =0$ (note that because $\mathcal{E}_6$ is a linear combination of $W^3_1$ and $W^3_2$ \cite{24}, the relation $\mathcal{E}_6=0$ becomes redundant), $W_{\mu\nu\alpha\beta}\nabla^{\nu} \nabla^{\beta}R^{\mu\alpha}=0$, $\nabla^\rho R^{\mu\nu\alpha\beta}\nabla_{\rho}W_{\mu\nu\alpha\beta}=0$, $R^{\mu\alpha}\nabla^{\beta}\nabla^{\nu}W_{\mu\nu\alpha\beta}=0$ and $\nabla^\mu W_{\mu\nu\alpha\beta}\nabla^{\alpha} R^{\nu\beta}=0$. Therefore, for conformally invariant space-times, there are 1 relations coming from the corollary of the Lovelock theorem, and 7 coming from $W_{\mu\nu\alpha\beta}=0$, so there are 8 scalars less in the reduced basis, i.e. 8 scalars left.

 \subsection{General order 6 linear combination for FLRW}
 
 The sum of all order six independent scalars, expressed in terms of $a(t)$ and its derivatives is :
 \begin{eqnarray*}
\begin{split} 
J=&\sum \Big( v_i \mathcal{L}_i +x_i \curv{L}_i \Big) =\frac{3}{a(t)^6} \Bigg( \sigma_1(v_i,x_i) \; \dot{a}(t)^6 + \sigma_2(v_i,x_i) \; a(t) \; \dot{a}(t)^4 \; \ddot{a}(t) + \sigma_3(v_i,x_i) \; a(t)^2 \; \dot{a}(t)^2 \; \ddot{a}(t)^2
   \\& + \sigma_4(v_i,x_i) \; a(t)^3 \; \ddot{a}(t)^3 + \sigma_5(v_i,x_i) \; a(t)^2 \; \dot{a}(t)^3 \; a^{(3)}(t)+ \sigma_6(v_i,x_i) \; a(t)^3 \; \dot{a}(t) \; \ddot{a}(t) \; a^{(3)}(t)
   \\& + \sigma_7(v_i,x_i) \; a(t)^4 \; a^{(3)}(t)^2+ \sigma_8(v_i,x_i) \; a(t)^3 \; \dot{a}(t)^2 \; a^{(4)}(t)+ \sigma_9(v_i,x_i) \; a(t)^4 \; \ddot{a}(t) \; a^{(4)}(t) \Bigg)
\end{split}
\end{eqnarray*}  
with,
 
$\left\{
\begin{array}{l}
\sigma _1=4 (2 v_1-2 v_3+6 v_4+2 v_5+2 v_6+6 v_7+18 v_8+2
   x_2+x_3+6 x_4\\
  \quad\quad -6 x_5-5 x_6-8 x_7-12 x_8)\vspace{6pt}\\
\sigma _2=2 \big(-2 v_3+12 v_4+4 v_5+6 v_6+24 v_7+108 v_8+30 x_1+x_2\\
\quad\quad -3 x_3-18 x_4+20 x_5+18 x_6+32 x_7+24 x_8\big)  \vspace{6pt}\
\\
\sigma _3= \big[ -6 v_2-4 v_3+24 v_4+14 v_5+6 v_6+48 v_7+216 v_8+48 x_1\\
\quad\quad +14 x_2+7 x_3+42 x_4-20 x_5-19 x_6-36 x_7-12 x_8 \big]\vspace{6pt}
\\ 
\sigma _4=8 v_1+2 v_2-8 v_3+24 v_4+6 v_5+10 v_6+24 v_7+72 v_8-12
   x_1-x_3-6 x_4\vspace{6pt}
\\  
\sigma _5=-2 \left(18 x_1+5 x_2+x_3+6 x_4-4 x_5-2 x_6-24 x_8\right)\vspace{6pt}
\\  
\sigma _6=-\left(36 x_1+8 x_2+x_3+6 x_4-2 x_6-8 x_7+24 x_8\right)\vspace{6pt}
\\ 
\sigma _7=-\left(4 x_5+3 x_6+4 \left(x_7+3 x_8\right)\right)  \vspace{6pt}
\\  
\sigma _8=-2 \left(6 x_1+x_2\right)\vspace{6pt}
\\  
\sigma _9=-\left(12 x_1+4 x_2+x_3+6 x_4\right)
\end{array}
\right.
$

Now in terms of $H(t)$ and its derivatives, considering only the scalars of the reduced FKWC basis : 
\begin{eqnarray*}
\begin{split} 
J=&\sum\limits_{i=4, 6, 7} v_i \mathcal{L}_i + \sum\limits_{j=1, 3, 5, 8} x_j \curv{L}_j  =3 \Bigg( \widetilde{\sigma}_1(v_i,x_i) \; H(t)^6 + \widetilde{\sigma}_2(v_i,x_i) \; H(t)^4 \dot{H}(t) + \widetilde{\sigma}_3(v_i,x_i) \; H(t)^2 \dot{H}(t)^2
   \\&  \quad \quad \quad \quad \quad \quad \quad \quad \quad \quad + \widetilde{\sigma}_4(v_i,x_i) \; \dot{H}(t)^3 + \widetilde{\sigma}_5(v_i,x_i) \; H(t)^3 \ddot{H}(t)+ \widetilde{\sigma}_6(v_i,x_i) \; H(t) \dot{H}(t) \ddot{H}(t)
   \\& \quad \quad \quad \quad \quad \quad \quad \quad \quad \quad + \widetilde{\sigma}_7(v_i,x_i) \ddot{H}(t)^2+ \widetilde{\sigma}_8(v_i,x_i) \; H(t)^2 H^{(3)}(t)+ \widetilde{\sigma}_9(v_i,x_i) \; \dot{H}(t) H^{(3)}(t) \Bigg)
\end{split}
\end{eqnarray*}  
with, 

$\left\{
\begin{array}{l}
\widetilde{\sigma} _1=12 \left(8 v_4+3 \left(v_6+4 v_7\right)\right)
\vspace{6pt}\\~
\widetilde{\sigma} _2=6 \left(24 v_4+9 v_6+36 v_7-48 x_1-2 x_3\right) 
\vspace{6pt}\\~
\widetilde{\sigma} _3= 4 \left(24 v_4+9 v_6+30 v_7-60 x_1-2 x_3-14 x_5-48
   x_8\right)
\vspace{6pt}\\~
\widetilde{\sigma} _4=2 \left(12 v_4+5 v_6+12 v_7-24 x_1-2 x_3\right)
\vspace{6pt}\\~
 \widetilde{\sigma} _5=-7 \left(24 x_1+x_3\right)
\vspace{6pt}\\~
\widetilde{\sigma} _6=-\left(84 x_1+5 x_3+24 \left(x_5+4 x_8\right)\right)
\vspace{6pt}\\~
\widetilde{\sigma} _7=-4 \left(x_5+3 x_8\right)
\vspace{6pt}\\~
 \widetilde{\sigma} _8=-\left(24 x_1+x_3\right)
\vspace{6pt}\\~
 \widetilde{\sigma} _9=-\left(12 x_1+x_3\right)
\end{array}
\right.
$

\subsection{$\nabla_\sigma R \nabla^\sigma R$ in Bianchi I spacetime}

We reproduce only here the value of the particular scalar $\nabla_\sigma R \nabla^\sigma R$ for Bianchi I spacetime, and the associated equations of motion of its square-root.
\begin{eqnarray}
\begin{split} 
~& \nabla_\sigma R \nabla^\sigma R= -4 \Bigg\{\alpha (t) \gamma (t) \dot{\beta} (t) \bigg(\alpha (t) \dot{\beta} (t) \dot{\gamma} (t)+\gamma (t)
   \big(\dot{\alpha} (t) \dot{\beta} (t)+\alpha (t) \ddot{\beta} (t)\big)\bigg)  +\beta (t)
   \bigg(\alpha (t)^2 \dot{\beta} (t) \dot{\gamma} (t)^2
   \\& 
 -\alpha (t)^2 \gamma (t) \big(\dot{\gamma} (t)
   \ddot{\beta} (t)+\dot{\beta} (t) \ddot{\gamma} (t)\bigg) +\gamma (t)^2 \bigg(\dot{\alpha} (t)^2 \dot{\beta} (t)-\alpha (t) \dot{\alpha} (t) \ddot{\beta} (t)-\alpha (t) \big(\dot{\beta} (t) \ddot{\alpha} (t)+\alpha (t) \beta ^{(3)}(t)\big)\big)\bigg)
   \\
   &+\beta (t)^2 \Bigg[\alpha (t)
   \dot{\gamma} (t) \bigg(\dot{\alpha} (t) \dot{\gamma} (t)+\alpha (t) \ddot{\gamma} (t)\bigg)+\gamma (t)^2
   \bigg(\dot{\alpha} (t) \ddot{\alpha} (t)-\alpha (t) \alpha ^{(3)}(t)\bigg)+\gamma (t)
   \bigg(\dot{\alpha} (t)^2 \dot{\gamma} (t)
   \\
   &-\alpha (t) \dot{\alpha} (t) \ddot{\gamma} (t)-\alpha (t)
   \Big(\dot{\gamma} (t) \ddot{\alpha} (t)+\alpha (t) \gamma
   ^{(3)}(t)\Big)\bigg)\Bigg]\Bigg\}^2 \, \Bigg/ \Bigg\{ \alpha (t)^4 \beta (t)^4 \gamma (t)^4 \Bigg\}
   \end{split} 
\end{eqnarray}  
The associated equations of motion, very complicated, yet second order, for respectively $\alpha (t) $, $ \beta (t) $ and $\gamma (t)$ are : 
\begin{eqnarray}
\begin{split} 
~&0 = -\alpha (t)^3 \gamma (t)^3 \dot{\beta} (t)^3+\alpha (t)^3 \beta (t) \gamma (t)^3 \dot{\beta} (t)
   \ddot{\beta} (t)+\alpha (t) \beta (t)^2 \gamma (t)^2 \Bigg[2 \alpha (t) \dot{\alpha} (t) \dot{\beta}(t) \dot{\gamma} (t)
   \\
   &+\gamma (t) \Big(\dot{\alpha} (t)^2 \dot{\beta} (t)+\alpha (t) \dot{\beta} (t)
   \ddot{\alpha} (t)+\alpha (t) \dot{\alpha} (t) \ddot{\beta} (t)\Big)\Bigg]+\beta (t)^3
   \Bigg[-\alpha (t)^3 \dot{\gamma} (t)^3+\gamma (t)^3 \Big(-2 \dot{\alpha} (t)^3
   \\
   &+3 \alpha (t)
   \dot{\alpha} (t) \ddot{\alpha} (t)\Big)+\alpha (t)^3 \gamma (t) \dot{\gamma} (t) \ddot{\gamma}(t)+\alpha (t) \gamma (t)^2 \Big(\dot{\alpha} (t)^2 \dot{\gamma} (t)+\alpha (t) \dot{\gamma} (t)
   \ddot{\alpha} (t)+\alpha (t) \dot{\alpha} (t) \ddot{\gamma} (t)\Big)\Bigg]
   \end{split} 
\end{eqnarray}  
\begin{eqnarray}
\begin{split} 
~&0 = -2 \alpha (t)^3 \gamma (t)^3 \dot{\beta} (t)^3+\alpha (t)^2 \beta (t) \gamma (t)^2 \dot{\beta} (t)
   \Bigg[\alpha (t) \dot{\beta} (t) \dot{\gamma} (t)+\gamma (t) \Big(\dot{\alpha} (t) \dot{\beta} (t)+3
   \alpha (t) \ddot{\beta} (t)\Big)\Bigg]
   \\
   &+\alpha (t)^2 \beta (t)^2 \gamma (t)^2
   \Bigg[\gamma (t) \dot{\beta} (t) \ddot{\alpha} (t)+\alpha (t) \dot{\gamma} (t) \ddot{\beta} (t)+\dot{\alpha}(t) \Big(2 \dot{\beta} (t) \dot{\gamma} (t)+\gamma (t) \ddot{\beta} (t)\Big)+\alpha (t) \dot{\beta} (t) \ddot{\gamma} (t)\Bigg]
   \\
   &+\beta (t)^3 \Bigg[-\alpha (t)^3 \dot{\gamma} (t)^3+\gamma (t)^3
   \Big(-\dot{\alpha} (t)^3+\alpha (t) \dot{\alpha} (t) \ddot{\alpha} (t)\Big)+\alpha (t)^3 \gamma
   (t) \dot{\gamma} (t) \ddot{\gamma} (t)\Bigg]
   \end{split} 
\end{eqnarray}  
\begin{eqnarray}
\begin{split} 
&0 = -\alpha (t)^3 \gamma (t)^3 \dot{\beta} (t)^3+\alpha (t)^3 \beta (t) \gamma (t)^3 \dot{\beta} (t)
   \ddot{\beta} (t)+\beta (t)^3 \Bigg[-2 \alpha (t)^3 \dot{\gamma} (t)^3\\
   &+\gamma (t)^3 \Big(-\dot{\alpha}(t)^3+\alpha (t) \dot{\alpha} (t) \ddot{\alpha} (t)\Big)
   \\
&+\alpha (t)^2 \gamma (t) \dot{\gamma} (t)
   \Big(\dot{\alpha} (t) \dot{\gamma} (t)+3 \alpha (t) \ddot{\gamma} (t)\Big)+\alpha (t)^2 \gamma
   (t)^2 \Big(\dot{\gamma} (t) \ddot{\alpha} (t)+\dot{\alpha} (t) \ddot{\gamma} (t)\Big)\Bigg]
   \\
   &+\alpha
   (t)^2 \beta (t)^2 \gamma (t) \Bigg[\alpha (t) \dot{\beta} (t) \dot{\gamma} (t)^2+\gamma (t)
   \Big(2 \dot{\alpha} (t) \dot{\beta} (t) \dot{\gamma} (t)+\alpha (t) \dot{\gamma} (t) \ddot{\beta} (t)+\alpha
   (t) \dot{\beta} (t) \ddot{\gamma} (t)\Big)\Bigg]
   \end{split} 
\end{eqnarray}

\bibliographystyle{mdpi}
\makeatletter
\renewcommand\@biblabel[1]{#1. }
\makeatother


\begin{thebibliography}{999} 


\bibitem{1}
Steven Weinberg.  The cosmological constant problem. {\em Rev. Mod. Phys.} {\bf 1989}, {\em  61}.

\bibitem{2}
Alexei A. Starobinsky.  A new type of isotropic cosmological models without singularity. {\em Physics Letters B.} {\bf 1980}, {\em  91}, 99-102.

\bibitem{3}
Robert H. Brandenberger.  A Nonsingular Universe. {\em Brown University.} {\bf 1992}.

\bibitem{4}
Woodard, R. Avoiding dark energy with 1/R modifications of gravity. {\em Lect.Notes Phys.} {\bf 2007}, {\em 720}, 403-433.


\bibitem{5} S. M. Carroll, V. Duvvuri, M. Trodden and M.
Turner. {\em Phys. Rev. D} {\bf 70} {\em 2004} 043528; S.
Capozziello, S. Carloni and A. Troisi. {\em Int. J. Mod. Phys D} {\bf 2003}, {\em 12} 1969.

\bibitem{6} 
 T.~P.~Sotiriou and V.~Faraoni.
  f(R) Theories Of Gravity.
  {\em Rev. Mod. Phys. }  {\bf 2010} {\em 82} 451.

\bibitem{7}
  S.~Nojiri and S.~D.~Odintsov.
  Unified cosmic history in modified gravity: from F(R) theory to Lorentz non-invariant models.
 {\em Phys. Rept.}  {\bf 2011}, {\em 505} 59.

\bibitem{8} 
  S.~Capozziello and M.~De Laurentis. Extended Theories of Gravity. {\em  Phys. Rept.}  {\bf 2011}, {\em 509} 167.


\bibitem{9}
Antonio De Felice, Shinji Tsujikawa. $f \big(R \big)$ theories. {\em   Living Rev. Rel.} {\bf 2010}, {\em 13: 3}.

\bibitem{10}
Lovelock, D. The uniqueness of the Einstein field equations in a four-dimensional space. {\em  Arch.Ration.Mech.Anal.} {\bf 1994}, {\em 33}, 54-70.

\bibitem{11}
C. Cherubini, D. Bini, S. Capozziello, R. Ruffini.Second Order Scalar Invariants of the Riemann Tensor: Applications to Black Hole space-times. {\em  Int.J.Mod.Phys.} {\bf 2002}, {\em D11}, 827-841.

\bibitem{12}
James T. Wheeler. Weyl gravity as general relativity. {\em  Phys. Rev.} {\bf 2014}, {\em D90}, 025027.

\bibitem{13}
Gregory Walter Horndeski. Second-order scalar-tensor field equations in a four-dimensional space. {\em Int. J. Theor. Phys.} {\bf 1974}, {\em 10}, 363-384.

\bibitem{14}
C. Charmousis, E. J. Copeland, A. Padilla, P. M. Saffin. General second order scalar-tensor theory, self tuning, and the Fab Four. {\em Phys. Rev. Lett.} {\bf 2012}, {\em 108}, 051101.

\bibitem{15}
S. Deser, O. Sarioglu, B. Tekin. Spherically symmetric solutions of Einstein + non-polynomial gravities. {\em Gen.Rel.Grav.} {\bf 2008}, {\em 40}, 1-7.

\bibitem{16}
Changjun Gao. Generalized modified gravity with the second-order acceleration equation. {\em Phys. Rev.} {\bf 2012}, {\em D86}, 103512.

\bibitem{17}
S A Fulling, R C King, B G Wybourne and C J Cummins. Normal forms for tensor polynomials. I. The Riemann tensor. {\em Class. Quantum Grav.} {\bf 1992}, {\em 9}, 1151.

\bibitem{18}
S. Deser, and A.V. Ryzhov. Curvature invariants of static spherically symmetric geometries. {\em Class.Quant.Grav.} {\bf 2005}, {\em 22}, 3315-3324.

\bibitem{19}
Ashtekar A. and Singh P.  Loop quantum cosmology: a status report. {\em Class. Quantum Grav.} {\bf 2011}, {\em 28}, 213001.

\bibitem{20}
P. Binetruy, C. Deffayet, U. Ellwanger and D. Langlois. Brane cosmological evolution in a bulk with cosmological constant. {\em Phys. Lett.} {\bf 2000}, {\em B477}, 285-291.

\bibitem{21} 
  G. Cognola, R. Myrzakulov, L. Sebastiani and S. Zerbini.
 Einstein gravity with Gauss-Bonnet entropic corrections.
  {\em Phys. Rev. D} {\bf 2013}, {\em 88}, 024006.

\bibitem{22} 
  G. Cognola, E. Elizalde, L. Sebastiani and S. Zerbini.
  Black hole and de Sitter solutions in a covariant renormalizable field theory of gravity.
  {\em Phys. Rev. D} {\bf 2011}, {\em 83}, 063003.

\bibitem{23}
E. Bellini, R. Di Criscienzo, L. Sebastiani and S. Zerbini. Black Hole entropy for two higher derivative theories of gravity. {\em Entropy} {\bf 2010}, {\em 12}, 2186.

\bibitem{24}
Julio Oliva, Sourya Ray. Classification of Six Derivative Lagrangians of Gravity and Static Spherically Symmetric Solutions. {\em Phys.Rev.} {\bf 2010}, {\em D82}, 124030.

\bibitem{25}
Yves D\'{e}canini, Antoine Folacci. FKWC-bases and geometrical identities for classical and quantum field theories in curved space-time. {\em Unpublished report.} {\bf 2008}.

\bibitem{26}
Adel Awad, Ahmed Farag Ali.  Planck-Scale Corrections to Friedmann Equation. {\em Central Eur.J.Phys.} {\bf 2014}, {\em 12}, 245-255.

\bibitem{27}
Pantelis S. Apostolopoulos, George Siopsis, Nikolaos Tetradis. Cosmology from an AdS Schwarzschild black hole via holography. {\em Phys.Rev.} {\bf 2009}, {\em 102}, 151301.

\bibitem{28}
Esra Russell, Can Battal Kilinc, Oktay K. Pashaev. Bianchi I Model: An Alternative Way To Model The Presentday Universe. {\em MNRAS.} {\bf 2014}, {\em 442},  2331-2341.

\bibitem{29}
Thomas Schucker, Andr\'{e} Tilquin, Galliano Valent. Bianchi I meets the Hubble diagram. {\em MNRAS.} {\bf 2014}, {\em 444},  2820.

\bibitem{30}
D. Lovelock. Divergence-free tensorial concomitants. {\em aequationes mathematicae} {\bf 1970}, {\em 4},  127-138.

\bibitem{31}
D. Lovelock.  Degenerate Lagrange densities involving geometric objects . {\em Arch. Ration. Mech. Anal.} {\bf 1970}, {\em 36},  293-304.




\end{thebibliography}


%

\end{document}